\documentclass[journal]{IEEEtran}

\def\allfiles{}
\usepackage{bm}
\usepackage{amssymb}
\usepackage{amsmath}
\usepackage{amsfonts}
\usepackage{graphicx}
\usepackage{subfigure}
\usepackage{cite}
\usepackage{enumerate}
\usepackage{url}
\usepackage{epstopdf}
\usepackage{float}
\usepackage{color}
\usepackage{multirow}
\usepackage{enumitem}
\usepackage{mathtools}
\usepackage{makecell} 
\usepackage[normalem]{ulem} 
\usepackage{array}

\makeatletter  
\newif\if@restonecol  
\makeatother

\usepackage[linesnumbered, lined, commentsnumbered, ruled]{algorithm2e}
\usepackage{algpseudocode}
\usepackage{algorithmicx}

\newcommand{\mm}{meta-material}
\newcommand{\mbs}{Meta-Backscatter}

\begin{document}

\title{Meta-Backscatter: A New ISAC Paradigm for Battery-Free Internet of Things}

\author{
\IEEEauthorblockN{
\normalsize{Xu~Liu},~\IEEEmembership{\normalsize Student~Member,~IEEE},
\normalsize{Hongliang~Zhang},~\IEEEmembership{\normalsize Member,~IEEE},
\normalsize{Kaigui~Bian},~\IEEEmembership{\normalsize Senior Member,~IEEE},
\normalsize{Xi~Weng},
and~\normalsize{Lingyang~Song},~\IEEEmembership{\normalsize Fellow,~IEEE}
}
\thanks{
Xu~Liu, Hongliang~Zhang, and Lingyang~Song are with the State Key Laboratory of Advanced Optical Communication Systems Networks, the School of Electronics, Peking University.

Kaigui~Bian is with the School of Computer Science, Peking University.

Xi~Weng is with the Guanghua School of Management, Peking University.
}}

\maketitle
\begin{abstract}
The {\mm} sensor has been regarded as a next-generation sensing technology for the battery-free Internet of Things (IoT) due to its battery-free characteristic and improved sensing performance.
The {\mm} sensors function as backscatter tags that change their reflection coefficients with the conditions of sensing targets such as temperature and gas concentration, allowing transceivers to perform sensing by analyzing the reflected signals from the sensors.
Simultaneously, the sensors also function as environmental scatterers, creating additional signal paths to enhance communication performance.
Therefore, the {\mm} sensor potentially provides a new paradigm of Integrated Sensing and Communication (ISAC) for the battery-free IoT system. 
In this article, we first propose a {\mbs} system that utilizes {\mm} sensors to achieve diverse sensing functionalities and improved communication performance.
We begin with the introduction of the {\mm} sensor and further elaborate on the {\mbs} system.
Subsequently, we present optimization strategies for {\mm} sensors, transmitters, and receivers to strike a balance between sensing and communication.
Furthermore, this article provides a case study of the system and examines the feasibility and trade-off through the simulation results. 
Finally, potential extensions of the system and their related research challenges are addressed.
\end{abstract}


\ifx\allfiles\undefined
\begin{document}
\fi

\section{Introduction}
\label{sec: introduction}
Internet of Things (IoT) is considered to be a pivotal enabler for advancing future production methods and lifestyles, particularly in terms of sensing~\cite{wikstrom_challenges_2020}.
To enable the next generation of sensing applications, such as health care and intelligent storage, it requires sensors to be battery-free and cost-efficient in ubiquitous deployment, which cannot be satisfied by commercial sensors~\cite{9482542}.
The {\mm} sensor is a promising solution for the battery-free IoT system due to the following characteristics.
To be specific, the {\mm} sensors consist of sub-wavelength meta-material units whose frequency responses, i.e., reflection coefficients, are sensitive to the conditions of sensing targets, such as temperature and gas concentration.
The sensors work as passive backscatter tags without delicate RF units, chips, or power supplies, and thus are notably battery-free and cost-efficient~\cite{liu_internet_2023}.  
Besides, the receivers can capture the frequency response of a {\mm} sensor through signal reflection and derive the sensing results from the frequency response.
Simultaneously, by functioning as environmental scatterers, these sensors hold the potential to provide additional paths for existing communication systems, thereby enhancing the communication performance.
Consequently, integrating this sensing functionality with communication systems is a win-win scheme.
In addition, the ISAC paradigm significantly reduces the extensive bandwidth and hardware requirements of {\mm} sensors, as those dedicated to communication systems can be reused.

\begin{figure}[!t]
  \centering
  \includegraphics[width=0.90\linewidth]{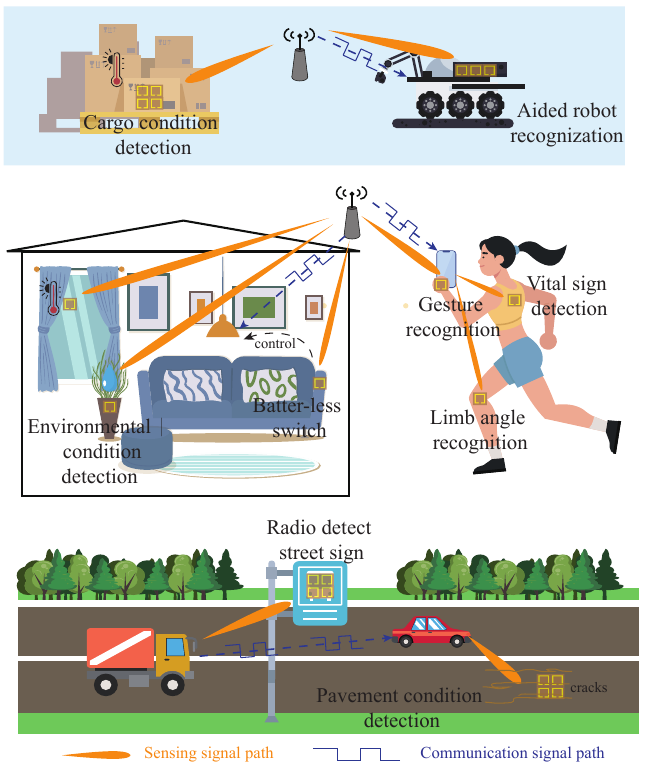}
  \vspace{-0.5em}
  \caption{{\mbs}: meta-material sensors enabled ISAC systems for the applications of the battery-free IoT system.}
  \label{fig: ISAC illustration}
  \vspace{-1.5em}
\end{figure}

\begin{table*}
\caption{Comparison of the Proposed ISAC with Existing Systems}
\label{tab: T2}
\centering
\begin{tabular}{m{3cm}<{\centering}m{6cm}<{\centering}m{6cm}<{\centering}}
\hline
\textbf{ } & \textbf{Proposed System}&\textbf{Existing ISAC Systems}\\
\hline
\textbf{Detected object} & Meta-material sensor & Reflector in environment \\
\hline
\textbf{Sensing functionality} & Diverse sensing functionalities & Only geometric and movement characteristics\\
\hline
\textbf{Working principle} &  Detecting the sensors' frequency response & Detecting the reflectors' reflection characters\\
\hline
\textbf{System objective} & \multicolumn{2}{c}{Sensing and communication are achieved through a single signal transmission}\\
\hline
\end{tabular}
\vspace{-2em}
\end{table*}

In this article, we introduce the {\mbs} system as a new ISAC paradigm for the battery-free IoT system, achieving diverse sensing with existing communication signals, as illustrated in Fig.~\ref{fig: ISAC illustration}.
The {\mbs} system consists of wireless transceivers and multiple {\mm} sensors that utilize communication signals for the sensing functionality.
Following the principle of backscatter propagation, the communication signals can show the frequency responses when reflected by the {\mm} sensors.
Through analyzing the reflected signals, the corresponding sensing results can be obtained by the receivers~\cite{hu_meta-material_2022}.
Simultaneously, the reflected signals also enhance the strength of the received signal to improve the communication performance between~transceivers.

In the literature, ISAC systems have attracted much attention from industry and academia due to the reduced spectral and spatial resources~\cite{liu_integrated_2022}.
However, current ISAC systems primarily focus on radar-based sensing of geometric and movement characteristics.
In contrast, the proposed system realizes diverse sensing functionalities for ISAC, including gas concentration, humidity distribution, vital signs of humans, and more, which can be detected with the assistance of {\mm} sensors.
The sensing functionality is achieved by detecting the reflected communication signals in both systems.
Specifically, the {\mm} sensors work as the reflectors and couple the information of the sensing targets on the reflected signal through their changeable frequency responses in the proposed system, bringing more possibilities and challenges.
The objective of both systems involves the reuse and integrated design to achieve sensing and communication through a single signal transmission.
To further clarify the proposed system, we present a comparison with existing ISAC systems in terms of the detected object, sensing functionality, working principle, and system objective, as shown in Tab.~\ref{tab: T2}.

As the {\mbs} system can accomplish sensing and communication at the same time, the cost of deployment and the occupation of the wireless spectrum are significantly decreased.
The proposed system can leverage existing wireless transceivers to achieve pervasive sensing based on ubiquitous {\mm} sensors and have the following potential applications as illustrated in Fig.~\ref{fig: ISAC illustration}.
\begin{itemize}[leftmargin=*]
	\item Smart home: By deploying {\mm} sensors throughout the building, it is possible to detect various environmental conditions. In addition, the sensor also can serve as a switch with multiple user-adjustable statuses.
	\item Human-computer interaction: By utilizing {\mm} sensors worn by individuals, the system can capture various human indicators, including gestures, limb movements, and body temperature, among others.
	\item Intelligent transportation: The {\mm} sensors have the potential for integration into street signs, transmitting information that can be read by RF devices in automobiles. Moreover, the sensors can be deployed on pavements to detect irregularities such as crevices. 
	\item Intelligent storage: The incorporation of {\mm} sensors into both robots and cargo enables the storage management system to continually monitor their status using communication signals.
\end{itemize}

However, an inherent trade-off between sensing and communication has existed in the {\mbs} system due to the integration between sensing and communication under limited system resources. 
Specifically, the sensing performance (i.e., accuracy and resolution) and communication performance (i.e., data rate) are mutually constrained under limited hardware, spectrum, and energy resources.
To realize the above visions, we employ a systematic approach that optimizes the sensor, transmitter, and receiver to strike the trade-off. 
More precisely, we provide the following contents.
\begin{itemize}[leftmargin=*]
	\item Meta-Backscatter system: We first illustrate the {\mm} sensor and further outline the {\mbs} system for the ISAC purpose.
	\item Key techniques: We clarify the system trade-off between sensing and communication and illustrate the optimization strategy on the structure of the {\mm} sensor, the waveform and beamforming of the transmitter, and the joint signal processing of the~receiver. 
	\item Simulation and evaluations: We present the {\mbs} system based on the OFDM signals as an instance, and the simulation results substantiate the feasibility of the system.
	\item Extensions and research challenges: We point out the potential extensions and corresponding research challenges of the {\mbs} system.
\end{itemize}

\ifx\allfiles\undefined
\end{document}
\fi

\ifx\allfiles\undefined
\begin{document}
\fi

\section{Basic of Meta-material Sensor and ISAC}
\label{sec: meta-material sensor}
In this section, we begin with the working principle of the {\mm} sensor.
Afterward, we introduce the {\mbs} system in which the {\mm} sensors are employed to improve the communication performance and provide the sensing functionality at the same time.

\subsection{Working Principle of Meta-material Sensors}
\label{ssec: sensor}
Meta-material sensors are passive wireless sensors that are constituted of meta-materials~\cite{liu_meta-material_2022}.
Meta-materials refer to artificial structures composed of sub-wavelength units and exhibit unique electromagnetic properties nonexistent in nature~\cite{weiglhofer2003introduction}.
These specific properties are reflected through the frequency response of the meta-material and determined by the electromagnetic and structure parameters of its components.
By designing the meta-material to have specific sensitive structures, the frequency response of the meta-material alters with the condition of the sensing target changes, for example, the temperature goes up.
The changeable frequency response causes the reflected signals to vary with the sensing target.
Therefore, the variations of the sensing target are detectable by receivers through analyzing the reflected signals from {\mm} sensors~\cite{liu_meta-material_2022}.

\begin{figure}[!t]
  \centering
  \includegraphics[width=0.80\linewidth]{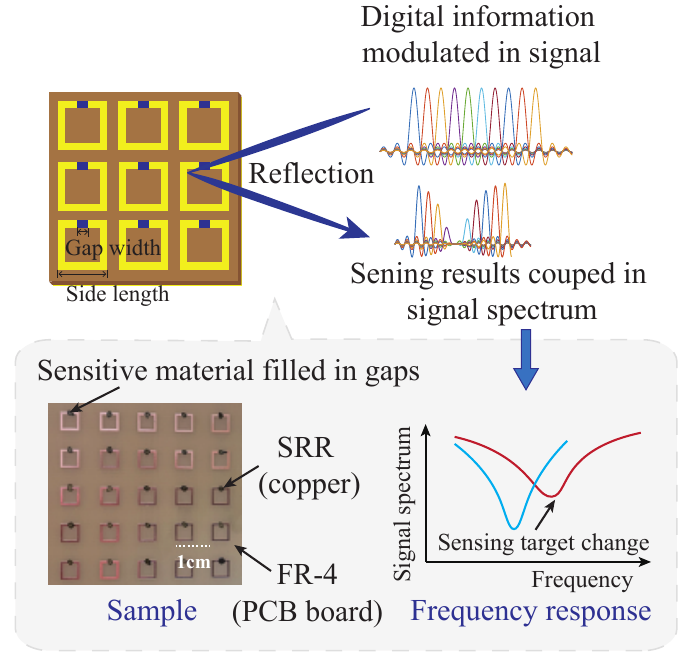}
  \vspace{-1em}
  \caption{The working principle and example of the {\mm} sensors.}
  \vspace{-1.5em}
  \label{fig: sensor}
\end{figure}

Fig.~\ref{fig: sensor} has shown an illustration to show an implementation of a {\mm} sensor utilizing a split-ring resonator (SRR) whose gap is filled with sensitive material (PEDOT-PSS) to the humidity.
Specifically, the resistance of the sensitive material changes with the humidity in the air to achieve humidity sensing.
A gap existing in the SRR provides resonance characteristics with specific frequency of electromagnetic waves~\cite{hu_meta-iot_2022}.
Based on the resonant characteristics provided by the SRR, the sensor exhibits a frequency-selected reflection that produces an absorption peak in the frequency domain of reflected signals.
Moreover, the resistance changes of the sensitive material influence both the position and shape of the absorption peak.
By recognizing these changes, the transceiver can effectively detect the humidity condition from the reflected signals.

Specifically, the sensor is fabricated using the printed circuit board (PCB) technique, which etches a two-dimensional metal pattern onto a substrate.
Thanks to this mature production technology, the production cost of the {\mm} sensors can be reduced to as low as \$0.1 per sensor.

The approach utilized by meta-material sensors resembles the principle of backscatter, which involves transmitting information by controlling the reflection of RF signals using a special device called a backscatter tag~\cite{xu_practical_2018}.
This similarity is the reason why we designate the proposed system as Meta-Backscatter.
However, the existing backscatter tags deploy an antenna to reflect and modify the signal through impedance mismatching, while meta-material sensors deploy sub-wavelength structures with changeable frequency response.
The sub-wavelength structures provide the concentrated reflected signals according to the phased array theory and high sensitivity towards the sensing target based on resonant characteristics, making the signal easily detectable by the transceiver and thus improving sensing performance~\cite{hu_meta-iot_2022}.

\subsection{Introduction of {\mbs} Systems}
\label{ssec: principle}

As introduced in the previous parts, the {\mm} sensors can function as environmental scatterers, and thus these sensors have the potential to provide additional paths for existing communication systems, meanwhile achieving the sensing functionality.
To achieve this win-win scheme, we propose the {\mbs} system.

\subsubsection{System Overview}
\begin{figure}[!t]
  \centering
  \includegraphics[width=0.90\linewidth]{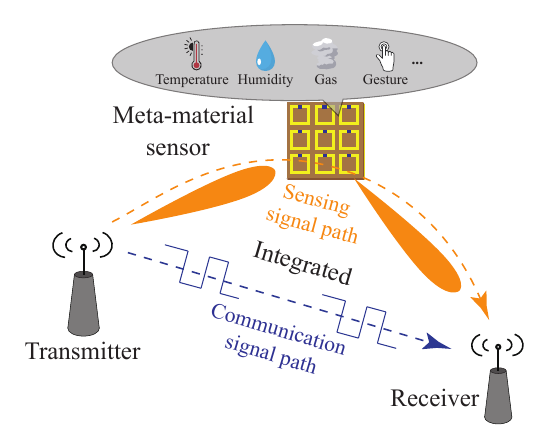}
  \vspace{-1.5em}
  \caption{The illustration of the {\mbs} system.}
  \label{fig: system}
  \vspace{-1.5em}
\end{figure}

To illustrate the technique clearly, we describe a basic system, which comprises one {\mm} sensor, one transmitter, and one receiver, as shown in Fig.~\ref{fig: system}. The large-scale system can be further discussed on this basis, such as multiple receivers or sensors.
The communication task has already existed between the transmitter and receiver based on wireless signals.
The transmitter such as base stations and access points emits signals that support both sensing and communication.
The receiver is equipped with a joint sensing and communication processing framework to analyze the signal.
The {\mm} sensor is deployed around to obtain the environmental conditions and incidentally provide additional paths for communication.
In the proposed system, the wireless signal transmits digital information and simultaneously works as a sensing signal in a single transmission.
It is worth mentioning that the proposed system can easily coexist with existing communication paradigms as the sensor can be simply regarded as the environmental scatterer.

As shown in Fig.~\ref{fig: sensor}, the {\mm} sensor can sense and transmit the sensing result by changing the frequency spectrum of wireless signals through reflection.
At the same time, signals reflected by {\mm} sensors can provide additional paths to improve communication performance. 
The ISAC procedures are elaborated as follows:

\begin{itemize}[leftmargin=*]
	\item For the sensing functionality, the receiver needs to extract the sensing results from the received signals. Two steps are involved to achieve sensing. First, since the meta-material sensor is an entirely passive device and cannot configure itself to a state that does not reflect the signal, there exists interference from other meta-material sensors and environmental scatterers. The transceiver must handle the interference, for example, through directional transmitting signals to specific sensors and signal processing establishing the relationship between the target and interference~\cite{liu_meta-material_2022}. Second, the frequency response of the sensor can be obtained by analyzing the average power spectral density of the received signal and comparing it with the power spectral density of the transmitted signal, which is known to the receiver. Then the sensing results can be determined by the obtained frequency response.
	\item For the communication functionality, the reflection of the {\mm} sensor creates an additional communication path between transceivers. Based on the dedicated sensor design, this new path can enhance the communication signal strength and further increase the system channel capacity, thereby improving the communication quality. However, the frequency selective absorption of the sensor causes spectrum distortion of the signal, leading to inter-symbol interference (ISI) during the communication, which must be handled through appropriate signal processing methods.
\end{itemize}

\subsubsection{Trade-off Between Sensing and Communication} 
Based on the deployment of the {\mm} sensor, the proposed system not only provides sensing functionality but also improves communication quality for users.
However, limited by finite system resources, there exists a trade-off between sensing and communication in this system.
\begin{itemize}[leftmargin=*]
	\item The sensing performance of the system is linked to variations in the signal arising from environmental conditions. The modulation of digital information on the signal amplifies the signal's inherent randomness, making it more challenging to detect variations resulting from the sensing target. Furthermore, the signal transmitted along the communication path reduces the visibility of absorption peaks, further heightening the sensing challenge.
	\item The communication performance of the system is determined by wireless channel conditions, such as the signal-to-noise ratio (SNR) during signal transmission. While the {\mm} sensors offer an additional communication path, the reflection also introduces frequency-selective fluctuations, leading to a partial loss of signal energy and a subsequent reduction in communication capacity.
\end{itemize}

In the following, we give discussions on how to achieve the balance between sensing and communication through well-designed {\mm} sensors, transmitters, and receivers in the proposed system.

\ifx\allfiles\undefined
\end{document}
\fi

\ifx\allfiles\undefined
\begin{document}
\fi

\section{Key Techniques for Meta-Backscatter System}
\label{sec: key_technique}

In this section, we discuss the trade-off between sensing and communication and propose the corresponding sensor, transmitter, and receiver design methods in this system.

\subsection{Structure Optimization for Meta-material Sensors}
\label{ssec: sensor optimization}
The structural parameters of the {\mm} sensor, such as the gap width, and side length of the SRR as shown in Fig.~\ref{fig: sensor}, determine its frequency response during the reflection of wireless signals~\cite{liu_meta-material_2022}.
The frequency response of the sensor affects both sensing and communication functionalities.
The desirable structural parameters must be chosen before the deployment of the sensor.

\textbf{Challenge}: To ensure both functionalities' performances, two targets related to the frequency response need to be considered in the structure optimization for the sensor.
\begin{itemize}[leftmargin=*]
	\item Sensing: The sensitivity depends on the variation in the frequency response with the sensing target, which should be maximized.
	\item Communication: The reflection efficiency determined by the frequency response of the {\mm} sensor must be optimized to ensure communication quality.
\end{itemize}

\textbf{Solution}: To balance these two objectives, we proposed the following schemes:

As demonstrated in Section~\ref{ssec: sensor optimization}, the frequency response of the sensor is presented as an absorption peak within the working frequency range.
The receivers recognize the position and shape of the absorption peak to determine sensing targets.
Therefore, the sensitivity of the sensor which determines the sensing accuracy, can be evaluated based on changes in the frequency response, including resonance frequency, Q-factor, and Euclidean distance of the absorption peak.
Specifically, an increase in Euclidean distance suggests more considerable changes in the absorption peak which is easier to detect and leads to more precise sensing outcomes.
It is essential to evaluate the sensing performance of the {\mm} sensor using these measurements to ensure its effectiveness.

However, the absorption peak leads to a decrease in signal strength, corresponding to a decrease in the SNR for communication.
Additionally, this frequency-selective reflection introduces channel frequency-selectivity, which can result in ISI in communication systems.
To guarantee the efficiency of communication, the frequency response should exhibit a flat response with reflection strength close to 100\% in the working frequency band~\cite{han_wideband_2019}.
However, this means the absorption peak has low Q-factors which increases the difficulty of detecting and further reduces the sensing accuracy of the system.
				 
In the optimization of sensor structures, the goal is to strike a balance between these two objectives. 
A multi-objective optimization problem is formulated to achieve a good performance trade-off between sensing accuracy and communication efficiency, facing different restrictions and demands in deployed environments~\cite{liu_internet_2023}.

\subsection{Beamforming and Waveform Optimization for Transmitters}
\label{ssec: transmitter optimization}
Since the wireless signal simultaneously accomplishes both sensing and communication tasks, transmitters must consider the joint waveform and beamforming for both functionalities to address the following conflicts caused by the meta-material sensor.
The sensing functionality relies on the signal reflection through the {\mm} sensor.
Meanwhile, the absorption of the sensor decreases the channel capacity for the communication functionality.
The waveform and beamforming must achieve a tradeoff between the two functionalities.
Without loss of generality, we assume that the transmitter is equipped with multiple antennas and works with compatible OFDM waveforms~\cite{cheng_hybrid_2021}.

\textbf{Challenge}: To ensure the sensing performance and maximize communication performance, the transmitter should adjust the waveform on subcarriers and the beamforming on antennas. These adjustments jointly determine the energy allocation between sensing and communication.
\begin{itemize}[leftmargin=*]
	\item Sensing: To ensure the sensing accuracy, the signal's strength at the resonance frequency must be guaranteed to resist the noise. Therefore, the energy allocation must compensate for energy loss caused by the absorption of sensors.
	\item Communication: To increase the communication capacity of the system, energy should be preferentially allocated to the subcarriers and line-of-sight (LoS) paths with low path losses. 
\end{itemize}

\textbf{Solution}: The waveform and beamforming can be dynamically adjusted at transceivers based on the following strategies.

Power allocation among subcarriers to balance the sensing and communication performance using the waveform method is a challenging task.
The {\mm} sensor absorbs the signal at specific frequencies, causing additional path loss on the corresponding subcarriers and resulting in lower channel gain.
From the communication perspective, allocating energy to subcarriers with higher channel gain corresponds to improved channel capacity.
From the sensing perspective, the absorption decreases the SNR of the signal, making the receiver struggle to distinguish the frequency response at these frequencies.
Thus, it is better to allocate energy to these subcarriers with lower channel gain to enhance the sensing accuracy.

Meanwhile, beamforming through multiple antennas needs to consider energy allocation between the sensor reflection path and other communication paths such as the LoS path.
Generally, the signal path reflected by the sensor introduces a higher path loss compared to the LoS paths without occlusion.
Therefore, the strategy of the beamformer focuses the energy on the LoS path to achieve a higher communication rate in the communication-only system.
Nevertheless, this strategy causes the excessively low strength of the reflected signal and increases the difficulty of detecting sensing signals.
To ensure the sensing accuracy, the beamformer must allocate energy to the sensor effectively, while still preserving the communication rate.

In practice, the system can dynamically adjust the parameters in different time slots to meet the different requirements of the users.

\subsection{Joint Sensing and Communication Process for Receivers}
\label{ssec: receiver optimization}
In the {\mbs} system, the receiver needs to process the received wireless signal to obtain both the digital communication information and sensing results.

\textbf{Challenge}: This dual-functional processing method faces these limitations.
\begin{itemize}[leftmargin=*]
	\item Sensing: The digital data modulated in wireless signal introduces uncertainty for the sensing estimation that significantly increases the difficulty of sensing detection.
	\item Communication: The frequency response of the sensor creates frequency-selective channels that result in ISI for the transmitted digital information.
\end{itemize}

As discussed in Section~\ref{ssec: sensor}, the {\mm} sensors use changes in the frequency response to transmit sensing information.
On the one hand, the irregular frequency response of the sensor introduces distortion in the waveform of the received signal, resulting in ISI and complicating the demodulation of the digital information.
On the other hand, the digital information in communication constrains the accuracy and temporal resolution of sensing.
The ISAC signal spectrum changes with not only the sensing target but also the modulation of digital information.
To handle the influence of the latter, the receiver needs to accumulate sufficiently long frames to acquire a fixed power spectral density, which limits the temporal resolution of the sensing.
Moreover, the sufficient accumulation of frames cannot always be satisfied, especially when facing rapidly changing sensing targets.
Subsequently, the acquired power spectral density of the received signal is not precise for detection, thereby decreasing the sensing accuracy.
To address these issues, it is imperative to develop a signal processing algorithm capable of extracting both the frequency response and the digital information.

\textbf{Solution}: The utilization of data-driven techniques based on deep learning for joint channel estimation and signal demodulation shows significant promise~\cite{qin_deep_2019}.
In communication systems, signal distortion, arising from changes in sensor frequency response, is primarily induced by the wireless transmission of signals through the environment.
Estimating the sensor frequency response equates to estimating channel parameters.
Deep learning-based signal processing algorithms can simultaneously address channel estimation and signal demodulation, which are handled independently in conventional receivers.
Additionally, the mapping of the extracted sensor frequency response to the obtained sensing results can be accomplished.
These data-driven approaches can handle complex channels such as the variational frequency response of the sensor to acquire better performance compared with traditional methods.

\ifx\allfiles\undefined
\end{document}
\fi

\ifx\allfiles\undefined
\begin{document}
\fi

\section{Simulation and Discussion}
\label{sec: example}
In this section, we provide a case study of using {\mm} sensors in an OFDM communication system that effectively supports broadband communication and resists frequency selective fading~\cite{xu2023dual}.
Firstly, we illustrate the simulation setup of the system.
Subsequently, we present the simulation results along with the discussions.

\subsection{Scenario Description}
Our research deals with a {\mbs} system, comprising one sensor, one base station, and one receiver as shown in Fig.~\ref{fig: system}~\cite{xu2023dual}.
The system works on the frequency range from $5.6\ \text{GHz}$ to $6.1\ \text{GHz}$.
The OFDM waveform is employed between the base station and receiver as an example to be compatible with existing communication systems.
The channel is modeled as the free space propagation and two signal paths including line-of-sight and the reflection of the meta-material sensor are considered.
We evaluate the communication performance of the system based on the average channel capacity and evaluate the sensing performance of the system based on the sensing accuracy of the measurement.
The sensing functionality infers the conditions of the sensing target from the received signals.
The discernibility of the reflected signal under different sensing target conditions determines the sensing performance.
The Euclidean distance of the reflected signal is one of the widely adopted discernibility measures, determining the sensing accuracy based on the maximum likelihood criterion~\cite{hu_meta-iot_2022}.

\subsection{System Optimization Strategy}
The research involves optimizing the sensor's structure and transmitter's waveform to balance the trade-off between sensing and communication in the system. 
Besides, this case study lacks research on the joint processing for the receiver, and the performance of dual functionalities is evaluated by average channel capacity and the Euclidean distance of the sensor reflected signal.
\subsubsection{Sensor optimization} The sensor provides an additional reflection path to enhance channel capacity.
Since the efficiency of the reflection depends on the structure of the sensor and determines the quality of the communication, we formulate a problem aimed at maximizing channel capacity through the adjustment of structural parameters.
Furthermore, as discussed in Section~\ref{ssec: sensor optimization}, the structure also influences sensing performance by determining the variation in the frequency response. 
Consequently, we incorporate sensing performance as a constraint in the optimization problem.
Specifically, the sensing accuracy is determined by the Euclidean distance of the reflected signals received under varying sensing target conditions.
Our algorithm guarantees system sensing performance by optimizing the sensor's structure, imposing a minimum sensing threshold on the average Euclidean distance, i.e., $\delta$.

\subsubsection{Waveform optimization} The communication signal loses a part of its energy and causes the decrease of the SNR due to the observation of the sensor.
Therefore, the system prioritizes allocating energy to frequencies with higher reflection efficiency in order to maximize channel capacity.
To ensure the sensing performance, this optimization problem also incorporates sensing performance constraints as part of the optimization problem.
In contrast, the optimization used in traditional power allocation schemes focuses on maximizing channel capacity while only considering signal power constraints~\cite {shen_fractional_2018}.

\subsection{Simulation Evaluation}

\begin{figure}[!t]
  \centering
  \includegraphics[width=0.85\linewidth]{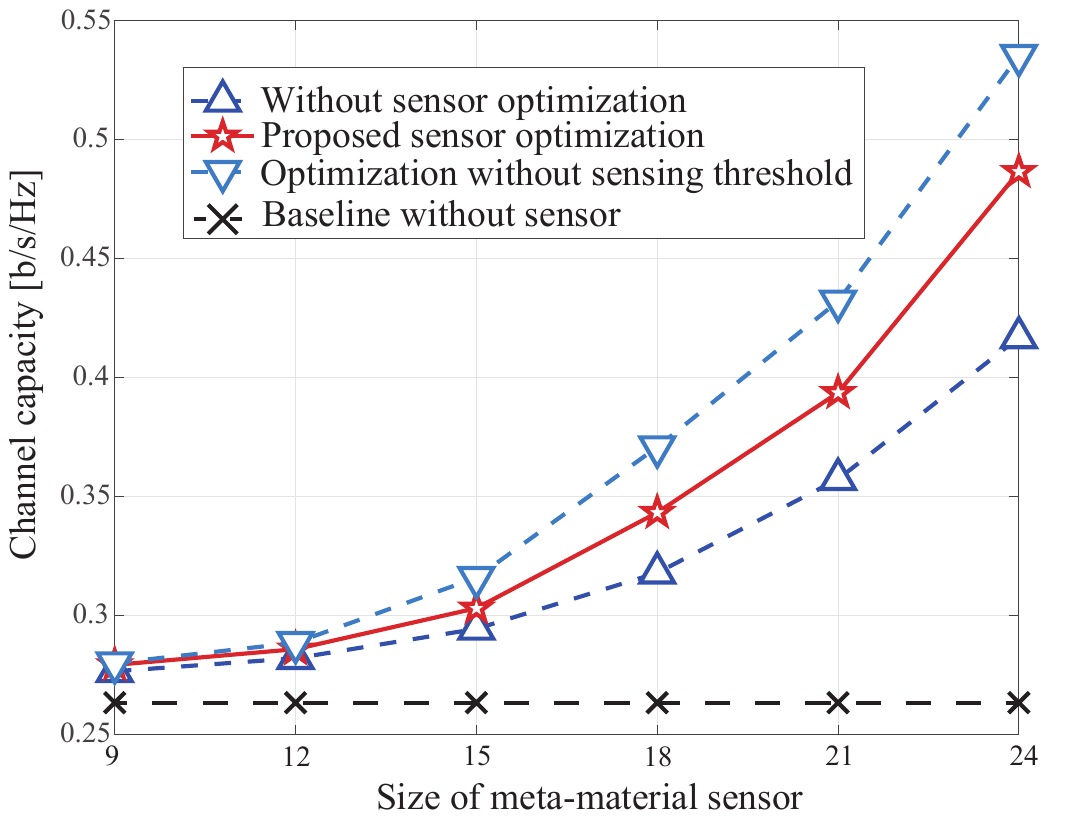}
  \vspace{-1em}
  \caption{The trade-off between sensing and communication exists in the sensor structure optimization, and the deployment of the {\mm} sensor improves the channel capacity.}
  \vspace{-1.5em}
  \label{fig: simulation result1}
\end{figure}

Fig.~\ref{fig: simulation result1} illustrates variations in average channel capacity with sensor sizes for different sensor structures.
The size of the {\mm} sensor is defined as the number of units on one side of the array.
The channel capacity is calculated with optimal power allocation for communication.
Three types of sensors with different structural parameters are used in the simulation.
The proposed scheme considers both sensing and communication functionalities.
The optimization without a sensing threshold does not consider the sensing constraint, which limits the minimum Eulerian distance.
The line without sensor optimization employs random sensor structural parameters.
Comparatively, it can be observed that under the proposed scheme, channel capacity decreases when compared to optimization without a sensing threshold.
Therefore, a trade-off between sensing and communication exists in sensor structure optimization.
Meanwhile, the baseline without sensors indicates the optimal channel capacity without the additional path provided by the {\mm} sensor.
The results demonstrate the effectiveness of the proposed scheme to increase the channel capacity through deploying the {\mm} sensor.
Moreover, although the communication capacity increases with the sensor size, the size of the sensor is limited by sensing.
As the distribution of the measured sensing target may not be uniform, the increased size results in inconsistent frequency response for each unit, misleading the receiver.

\begin{figure}[!t]
  \centering
  \includegraphics[width=0.87\linewidth]{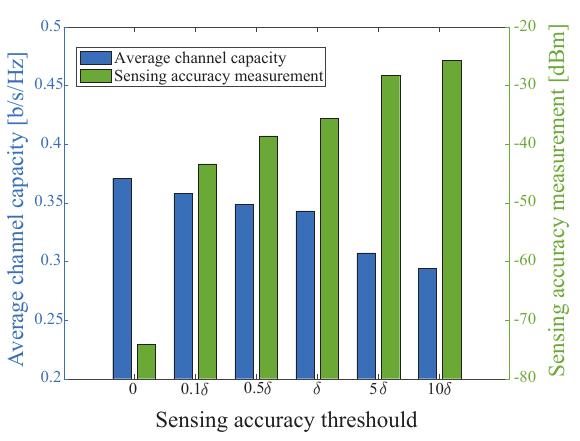}
  \vspace{-1em}
  \caption{The trade-off between sensing and communication exists in the {\mbs} system.}
  \vspace{-2em}
  \label{fig: simulation result2}
\end{figure}

Fig.~\ref{fig: simulation result2} presents the changes in the average channel capacity and sensing accuracy under different sensing constraints.
The sensing accuracy is presented through the actual signal power change of the received signal.
The $\delta$ is defined as the Euclidean distance in the initial average power allocation of the system
It can be observed that the average channel capacity drops as the threshold grows while the sensing accuracy increases. 
Therefore the trade-off between sensing and communication exists in the proposed system.
Furthermore, the reduction in communication performance due to the sensing accuracy threshold is justified, as the waveform tends to allocate energy to the observed frequency to ensure the Euclidean distance.

\ifx\allfiles\undefined
\end{document}
\fi

\ifx\allfiles\undefined
\begin{document}
\fi

\section{Extensions and Future Outlook}
\label{sec: future}

In previous sections, we have illustrated the working principle of the {\mbs} system and demonstrated its feasibility.
In this section, we present the corresponding research challenges and potential extensions of the proposed system.

\subsection{Fundamental Trade-off Between Dual Functionalities}
Meta-Backscatter system integrates both sensing and communication, inevitably resulting in a trade-off between these two functionalities.
In different scenarios, the system may prioritize one objective over the other, leading to a trade-off.
Consequently, it is crucial to establish a theory that comprehensively describes the inherent trade-off between sensing and communication within this system in a quantitative and unified manner.
The main challenge is that the major performance metrics and bounds used in sensing, communications, and ISAC need to be explicitly defined.
We only evaluate the two functionalities in limited aspects, i.e., sensing accuracy and channel capacity, based on the detection and information theories.
An achievable joint metric and its upbound are required to provide a general analysis framework for the Meta-Backscatter system.

\subsection{Large-scale Meta-Backscatter Systems}
To achieve ubiquitous sensing, the deployment of a large-scale system with multiple sensors and transceivers is a critical task in practice.
In such cases, various innovative sensing tasks, such as measuring environmental condition distributions, may be accomplished by integrating information from multiple {\mm} sensors~\cite{liu_meta-material_2022}.
Furthermore, multiple transceivers also can jointly sense a single meta-material sensor to resist interference from multi-paths and obstacles.
However, the information exchange and cooperative sensing processes have yet to be investigated.
It is worth noting that the exchange of sensing information consumes communication resources, introducing a new tradeoff between improved sensing performance and the occupation of communication resources. 
Additionally, the interference and integration among multiple sensors also pose under-explored research challenges.

\subsection{Terahertz Waves of Meta-Backscatter}
Intuitively, the higher carrier frequency can offer abundant bandwidth and increased sensing resolution, as observed in Terahertz (THz)~\cite{chaccour_seven_2022}. 
For instance, the narrow beams generated by the antenna array at THz frequencies can be aligned with the sensor to reduce interference from other ambient scatterers. 
However, the high path loss associated with THz signals restricts the coverage area for each transceiver. 
Consequently, the system relies on the intensive deployment and cooperation of transceivers in the network. 
A novel network topology such as cell-free design methodology is required to support efficient cooperation between transceivers for communication and sensing functionalities.

\ifx\allfiles\undefined
\end{document}
\fi

\section{Conclusion}
\label{sec: conclusion}
In this article, we have proposed a new ISAC paradigm based on {\mm} sensors to realize a general sensing functionality for the battery-free IoT system.
We have described the {\mbs} system and clarified the working principle of the system.
After that, we demonstrated the trade-off between sensing and communication that existed in the system.
Besides, the key techniques deployed to balance both functions have been proposed.
Furthermore, we have introduced an instance of using {\mm} sensors in OFDM communication and illustrated the feasibility and the trade-off in the system.
As a final prospect, we have provided the potential extensions and corresponding challenges of the {\mbs} system.

\bibliographystyle{IEEEtran}
\bibliography{bibilio}

\section*{Acknowledgments}
This work was supported in part by the National Key R\&D Project of China under Grant No. 2022YFE0111900; in part by the National Science Foundation under Grants 62371011, 62227809, and 62271012; and in part by the Beijing Natural Science Foundation under Grants L212027 and 4222005.

\section*{Biographies}

\vspace{-12mm}
\begin{IEEEbiographynophoto}{Xu Liu}  (xu.liu@pku.edu.cn) received a B.S. degree from the College of Engineering, Peking University, in 2021. He is currently pursuing a Ph.D. degree with the School of Electronics, Peking University.
\end{IEEEbiographynophoto}

\vspace{-12mm}
\begin{IEEEbiographynophoto}{Hongliang Zhang}  (hongliang.zhang@pku.edu.cn) is an assistant professor at the School of Electronics, Peking University. He was the recipient of the 2021 IEEE Comsoc Heinrich Hertz Award.
\end{IEEEbiographynophoto}

\vspace{-12mm}
\begin{IEEEbiographynophoto}{Kaigui Bian}  (bkg@pku.edu.cn) is an associate professor at the School of Computer Science, Peking University. His research interests include wireless networking and mobile computing.
\end{IEEEbiographynophoto}

\vspace{-12mm}
\begin{IEEEbiographynophoto}{Xi Weng}  (wengxi125@gsm.pku.edu.cn) is a professor at the Guanghua School of Management, Peking University. His research interests include game theory.
\end{IEEEbiographynophoto}

\vspace{-12mm}
\begin{IEEEbiographynophoto}{Lingyang Song} (lingyang.song@pku.edu.cn) joined the School of Electronics, Peking University in May 2009, where he is currently a Boya Distinguished Professor. He was a recipient of the IEEE Leonard G. Abraham Prize in 2016 and the IEEE Heinrich Hertz Award in 2021.
\end{IEEEbiographynophoto}

\end{document}